
\input harvmac
\def\np#1#2#3{Nucl. Phys. B{#1} (#2) #3}
\def\nc#1#2#3{Nuovo Cim. {#1} (#2) #3}
\def\pl#1#2#3{Phys. Lett. {#1}B (#2) #3}

\def\physrev#1#2#3{Phys. Rev. {D#1} (#2) #3}

\def\prep#1#2#3{Phys. Rep. {#1} (#2) #3}

\def\pf{{\rm Pf ~}}

\def\mpl{m_{{pl}}}
\def\al{\alpha}
\def\bt{\beta}

\lref\tHooft{G.~'t Hooft, in {\it Cargese Summer Inst. on Recent
Developments in Gauge Theories},  Cargese, France, 1979.}
\lref\Yanagida{T.~Hotta, K.-I.~Izawa and T.~Yanagida, {\it Dynamical Models
for Light Higgs Doublets in Supersymmetric Grand Unified Theories},
hep-ph/9509201.}
\lref\Maekawa{N.~Maekawa, {\it Duality of a Supersymmetric Standard
Model}, KUNS-1361,  hep-ph/ 9509407.}
\lref\VJreview{R.R.~Volkas and G.C.~Joshi, \prep{78}{1987}{1437}.}
\lref\DNNS{M.~Dine, A.~E.~Nelson, Y.~Nir and Y.~Shirman, {\it New Tools
For Low-Energy Dynamical Supersymmetry Breaking}, SCIPP-95-32,
hep-ph/9507378.}
\lref\CERN{K.~Konishi and K.~Shizuya, \nc{90A}{1985}{111};
D.~Amati, K.~Konishi, Y.~Meurice, G.C.~Ross and G.~Veneziano,
\prep{162}{1988}{169}.}
\lref\seiberg{N.~Seiberg, \physrev{49}{1994}{6857}, hep-th/9402044.}%
\lref\sem{N.~Seiberg, \np{435}{1995}{129}, hep-th/9411149.}%
\lref\kipp{K. Intriligator and P. Pouliot,
\pl{353}{1995}{471}, hep-th/9505006.}
\lref\berkooz{M. Berkooz, {\it The Dual of Supersymmetric SU(2k)
with an Antisymmetric Tensor and Composite Dualities}, RU-95-29,
hep-th/9505088.}
\lref\ilst{K.~Intriligator, R.G.~Leigh, M.J.~Strassler, {\it New Examples of
Duality in Chiral and Non-chiral Supersymmetric Gauge Theories},
RU-95-38, hep-th/9506148, to appear in Nucl.~Phys. B.}
\lref\kinsrev{K. Intriligator and N.~Seiberg, {\it Lectures on
 Supersymmetric Gauge Theories and Electric-Magnetic Duality},
RU-95-48, hep-th/9509066.}
\lref\barger{see, for example, V.~Barger, M.S.~Berger, P.~Ohmann,
\physrev{47}{1993}{1093}; \physrev{47}{1993}{2038}.}
\lref\sswitt{E.~Witten, \pl{105}{1981}{267}.}
\lref\ssnem{D.~Nemeschansky, \np{234}{1984}{379}.}
\lref\ann{A.E.~Nelson, unpublished.}

\Title{hep-ph/9510342, RU-95-69}
{\vbox{\centerline{Generating a Fermion Mass Hierarchy in a}
\centerline{}
\centerline{Composite Supersymmetric Standard Model}
}}

\centerline{M.J. Strassler}
\vglue .5cm
\centerline{Department of Physics and Astronomy}
\centerline{Rutgers University}
\centerline{Piscataway, NJ 08855-0849, USA}

\bigskip

\noindent

A mechanism is suggested by which the dynamics of confinement could be
responsible for the fermion mass matrix.    In this approach the large
top quark Yukawa coupling is generated naturally during confinement, while
those of the other quarks and leptons stem from non-renormalizable couplings
at the Planck scale and are suppressed.  Below the confinement scale(s)
the effective theory is minimal supersymmetric $SU(5)$ or the
supersymmetric standard model.  Particles in the $\bar 5$
representations of $SU(5)$ are fundamental while those in the $10$
and $5$ are composite.   The standard
model gauge group is weakly coupled and predictions of unification can be
preserved.    A  hierarchy in confinement
scales helps generate a hierarchical spectrum of quark and lepton
masses and ensures the Kobayashi-Maskawa matrix is nearly diagonal.
However, the most natural outcome is that the strange quark is
heavier than the charm quark; additional structure is required to
evade this conclusion.  No attempt has been made to
address the issues of $SU(5)$ breaking, SUSY breaking,
doublet/triplet splitting or the $\mu$ parameter.  While the models
presented here are neither elegant nor complete, they are remarkable
in that they can be analyzed without uncontrollable dynamical assumptions.

\Date{10/95}

Many authors over the years have proposed scenarios in which
some or all of the particles of the standard model are composite.
The work of 't Hooft \refs\tHooft\ on anomaly matching provided some
important consistency conditions on compositeness, but most
approaches have been limited by the need to make assumptions about
strongly coupled gauge theories.  Within the context of supersymmetric
model building, the idea that quarks and leptons might be the
supersymmetric partners of composite pseudo-Nambu-Goldstone bosons was
studied in detail during the 1980's;  Ref.~\refs\VJreview\ provides a
review of the extensive literature.

In the last couple of years, our understanding of the non-perturbative
dynamics of supersymmetric gauge theories has improved.  The new
methods are very powerful and can be used
for model building. In \refs\Yanagida\ a variant of the missing partner
mechanism was shown to be valid even at strong coupling.  In
\refs\Maekawa, following \refs\sem, it was suggested that the $SU(3)$
color group was the dual of another $SU(3)$, which implies
magnetic quarks and numerous composite Higgs doublets; unfortunately
it also implies $\Lambda_{QCD}\gg m_Z$.

In this letter I propose to use the dynamics of confinement as discussed in
\refs\seiberg\ to generate the fermion mass hierarchy in a supersymmetric
version of the standard model.\footnote{*}{When this paper was
complete I learned that this mechanism was previously
studied by A.~Nelson \refs{\ann}.}  Because
of recent developments involving duality of N=1 supersymmetric
theories (see \refs\kinsrev\ for a review) the results of \refs\seiberg\
now satisfy a large number of consistency checks far
beyond simple matching of global anomalies.  In this sense the
scenario presented here is free of uncontrollable dynamical assumptions.

It is convenient to present the new mechanism within the context of
minimal $SU(5)$ supersymmetric grand unified theories, though it
can also be applied directly to the standard model gauge group.
I will first present a toy one-generation model in which the basic physics is
explained; I will show that an up-type Yukawa coupling can be
large, while down-type Yukawa couplings are naturally smaller.  From
there I turn to theories with three generations.  The attractive features
of the mechanism are displayed in a simple model.  The top quark is
naturally heavy, the splittings of up-type quark masses are
generically larger than those of down-type quark masses, and
the Kobayashi-Maskawa matrix is naturally close to unity.   Unfortunately,
the charm quark is naturally too light relative to the strange quark and
additional physics must be invoked to avoid this conclusion.  I
present two approaches to achieving a reasonable mass spectrum, though
neither is especially elegant.  However, many other variants of these models
can be constructed.  An important
unresolved issue involves the doublet-triplet splitting problem, for
which I have presented no solution here. I have also left unaddressed
the questions of $SU(5)$ breaking and supersymmetry breaking.
Despite the weaknesses of these models, I hope that the reader
will find them amusing and thought-provoking.

\newsec{A toy model with one generation}

The main issue is to generate a $10$ representation of $SU(5)$ with a
coupling to the Higgs boson in the $5$ representation.  Antisymmetric
tensors of $SU(N)$ can be generated via the Berkooz trick
\refs{\berkooz,\ilst}, using an $SU(N)\times Sp(M)$ model, in which they
emerge as bound states of two $Sp(M)$ quarks; the $Sp$ group
confines for appropriate choice of $M$ and only the $SU(N)$
group remains at low energies.  A non-perturbative
superpotential is also generated \refs{\sem,\kipp,\berkooz};
we will see that for $SU(5)\times SU(2)$ (recall $Sp(1)\approx SU(2)$)
that this corresponds to the top-quark Yukawa coupling.  The
bottom-quark Yukawa coupling will have its source in a
non-renormalizable operator of dimension four\footnote{*}{Here
operator dimensions are given as appropriate for the superpotential,
{\it not} the Lagrangian; thus the bottom-quark Yukawa comes from an
quartic operator in the superpotential which is
dimension five in the Lagrangian.}
 and will be suppressed.

At the Planck scale\footnote{**}{Throughout
this letter I write $\mpl$ to signify whatever scale is the appropriate
ultraviolet boundary condition at which all non-renormalizable
couplings are generated at order one.  This may be $\mpl$ or $\mpl/4\pi$
or the string scale; the general mechanism presented here is insensitive
to the exact value of the scale.} $\mpl$  consider a theory with gauge group
$SU(5)\times SU(2)$.  The coupling of $SU(5)$ is weak while that of $SU(2)$
is large and blows up at the confinement scale $\Lambda_c\equiv \eta\mpl$.
The matter content of the theory, in terms of
$SU(5)\times SU(2)$ multiplets, consists of a field $X$ in the
$(5,2)$ representation, a field $S$ in the $(1,2)$, two fields
$ \bar H_i$ in the $(\bar5,1)$, and a massive field $\phi$
in the adjoint of $SU(5)$ whose sole purpose is to break
$SU(5)$ to the standard model.
All couplings in the superpotential which are consistent with the symmetries
are assumed to be present in the effective Lagrangian; all dimensions
are assumed to be $\mpl$ and all dimensionless coefficients are order one.
The lowest-dimension gauge-invariant operators which do not
involve $\phi$ are $\bar H_i X S$,  $\bar H_1 XX\bar H_2$, $X^5S$,
$(\bar H_iXS)(\bar H_jXS)$, {\it etc.}
Additional powers of $\phi$ can always be inserted to make new operators.
(The fact that this model breaks supersymmetry \refs\DNNS\
when $SU(5)$ becomes strongly coupled is irrelevant for the analysis
near $\Lambda_c$; models with three generations will not break
supersymmetry by themselves.)

Dynamics of strong coupling \refs{\seiberg} will drive
the $SU(2)$ group (which has six doublets) to confine near the
scale $\Lambda_c$.  Below this scale, $SU(2)$ with six doublets $q_i$
has {\it composite massless degrees of freedom}: the mesons
$V_{ij}=q_i q_j$, which are
antisymmetric in flavor \refs{\CERN,\seiberg}.   Classically this field
satisfies the constraint $\partial (\pf V )/ \partial V_{ij}=0$; quantum
mechanically the constraint is unmodified and is implemented by the
superpotential $W= \Lambda_c^{-3}\pf V$ \refs\seiberg.

In this model, the $SU(2)$ dynamics can be analyzed similarly; the
$SU(5)$, whether broken or not at the scale $\Lambda_c$,
is weakly coupled and is a spectator to the confining dynamics
of $SU(2)$.  The six doublets of $SU(2)$ consist of five from $X$
and one from $S$; after confinement the low-energy massless fields are
$A^{[\al\bt]}= X^\al X^\bt/\Lambda_c$ in the $10$ of $SU(5)$ and
$H^\al= X^\al S/\Lambda_c$ in the $5$ of $SU(5)$, where $\al,\bt$
are $SU(5)$ indices.  (I have defined $A$ and $H$ as
canonically normalized fields.)  As above a superpotential is generated
of the form
\eqn\Pfaff{
W_L =  (X^5S)\Lambda_2^{-3} = A A H
}
This term has the structure of the top-quark--Higgs-boson Yukawa coupling.
The coefficient of this term is order one; additional
contributions to this coefficient from tree-level dimension-six
terms at the Planck scale will be be suppressed by $\eta^3$.

Let us now analyze the theory below the scales of the strong dynamics.
For the moment, and for simplicity only, let us assume that
$M_{GUT}<\Lambda_c=\eta\mpl$, so that $SU(5)$ is unbroken at
$\Lambda_c$.\footnote{*}{An interesting challenge would be to
relate the scale $M_{GUT}$ to the dynamical scale $\Lambda_c$,
though as yet I know of no specific mechanism for doing so.}
The low energy theory consists of the fields $A,H,\bar H_i, \phi$
in the $10,5,\bar5,24$ of $SU(5)$ --- a standard model generation,
along with a pair of up- and down-type Higgs multiplets.
The light fields are coupled by the renormalizable superpotential
\eqn\toyW{
W =  m\bar H H +  Y A A H + y A \bar H \bar Q
}
where I have defined $\bar H$ as the linear combination of the $\bar H_i$
which couples to $H$, and $\bar Q$ as the orthogonal combination.
Of course there are many higher dimension terms.  The mass $m$ is
of order $\Lambda_c$; if we want to forbid it we may do
so by adding a discrete gauge symmetry under which $\bar H$, $\bar Q$
change sign.  The coupling $Y$, which was generated dynamically, is
order one.  The coupling $y$, which stems from a dimension-four term
${1\over\mpl}\bar H_1 XX\bar H_2$ in the tree-level superpotential at $\mpl$,
is of order $\eta$.   After $\phi$ condenses at
the scale $M_{GUT}$, the theory has a single standard model generation;
the top quark Yukawa coupling is order one, while the bottom quark and
tau lepton couplings are order $\eta$.  (Recall that the actual masses of
the top and bottom quarks involve the additional parameter
$\vev{H}/\vev{\bar H}\equiv \tan\bt$, so their ratio is not predicted.)

It is essential to note that this result depends crucially on the
choice of gauge group.  A top quark coupling of order one can only be
generated in this way if the weak gauge group is $SU(5)$ or one if
its subgroups and if the confining gauge group is $SU(2)$.  Of
course the mechanism can be embedded in a larger gauge group
which breaks at some scale to $SU(5)\times SU(2)$; thus we
can use $SU(k)\times Sp(m)$ for $k>5$, for example.

The confinement must take place at energies near the Planck scale;
otherwise the dimension-four operator which becomes the bottom-quark
Yukawa coupling after confinement will have too small a coefficient.
However, it need not be that $M_{GUT}<\Lambda_c$.
As we take $\Lambda_c$ equal to or smaller than $M_{GUT}$, the
low-energy model changes little, since the analysis of the superpotential
and of the size of its couplings is insensitive to the breaking of the
perturbative $SU(5)$ gauge group.  For this reason the analysis
will also work if the weakly coupled gauge group is $SU(3)\times SU(2)
\times U(1)$.

\newsec{A  model with three generations}

Next, I turn to a three-generation model.
The simplest implementation involves replicating the previous
structure three times.  At the Planck scale $\mpl$, the theory has
gauge group $SU(5)\times [SU(2)]^3$, where the gauge coupling of
the first factor is small while those of the last three are larger and
diverge at the scale $\Lambda_c$.

The matter content of the theory, in terms of
$SU(5)\times SU(2)\times SU(2)\times SU(2)$ multiplets, is as follows.
There are three fields $X_1,X_2,X_3$ which are in the representations
$(5,2,1,1), (5,1,2,1)$ and $(5,1,1,2)$ respectively.
There are fields $S_1,S_2,S_3$ in the $(1,2,1,1), (1,1,2,1)$
and $(1,1,1,2)$, and fields $\bar H_i, \bar Q_i$, $i=1,2,3$,
in the $(\bar 5,1,1,1)$.   There is also a field $\phi$
in the adjoint of $SU(5)$ which breaks $SU(5)$ to the
standard model.  Avoiding proton decay and keeping the Higgs bosons light
requires a gauged discrete symmetry, which I take to be a $Z_6$ under
which the fields $X_i,S_i,\bar H_i,\bar Q_i,\phi$ have charge $1,1,1,3,0$.
(This is anomaly-free under $SU(5)\times [SU(2)]^3$; instantons of
each group leave it unbroken.)  The lowest-dimension gauge-invariant
operators which can appear in the
superpotential (and do not depend on $\phi$) are
$\bar H_iX_jX_j\bar Q_k$,
$[(X_i^2)(X_j^2)(X_k S_k)]$, $[(\bar H_iX_jS_j)^2]$, {\it etc.}

Each $SU(2)$ subgroup confines, generating fields $A_i$ and
$H_i$ in the $10$ and $5$ representations of $SU(5)$ and
a superpotential $W = A_iA_iH_i$ with coefficient of order one.
In terms of $SU(5)$ representations,  the low energy theory
consists of  three standard model
generations along with three pairs of up- and down-type
Higgs doublets and their color-triplet partners.
The light fields are coupled by the renormalizable superpotential
\eqn\GUTW{
W = Y^{ijk} A_i A_j H_k+ y^{ijk} A_i \bar Q_j \bar H_k
}
where all repeated indices are summed.
Of course there are many non-renormalizable terms.
The structure of these terms is as before: $Y^{iii}$ is of
order one, all other $Y^{ijk}$ are of order $\eta^3$, and
the $y^{ijk}$ are all of order $\eta$.

However, since all three pairs of Higgs bosons are light, and
since all up-type Yukawa couplings are large, one must
explain why only one up-type Higgs boson gets a large vacuum expectation
value, why it couples mostly to the top quark, and why the vacuum
expectation value of the
down-type Higgs bosons couples mostly to the bottom quark.  Without treating
the third generation differently from the other two, this seems
challenging at best.

\newsec{A more realistic and less flavor-symmetric model.}

If one does treat the third generation differently,
the number of Higgs boson pairs can be reduced to one.  This can easily
be done, at the cost of simplicity, by altering the discrete symmetries.
One should take care, however, that $V_{tb}$ be close to unity.
This is not trivial to guarantee.  Suppose we permit $H_1, H_2,
\bar H_1,\bar H_2$ to become massive as a result of a discrete symmetry,
while $H_3$ and $\bar H_3$ remain light.  We have ensured dynamically
that $H_3$ couples only to $A_3$ at leading order,
so we may identify $A_3$ as containing the left-handed top and bottom quarks.
The mass matrix of the down-type quarks is $y^{ij3}$; we may use the
$SU(3)$ acting on $\bar Q_i$ (assuming it is not broken by the
discrete symmetry) to set $y^{313}$ and $y^{323}$ to zero.
To ensure that the bottom-quark is the most massive down-type
quark and that $V_{tb}$ is order one, we must have $y^{333}$ much
larger than any other element of the matrix.  But no symmetry
guarantees this, and since the couplings $y$ were
assumed to be generated at the Planck scale, they are in fact
naturally all of the same order.  Fortunately, a hierarchy in
confinement scales will assure this automatically.

One path to a reasonable model, treating the third
generation differently from the others, is to change the gauge symmetry.
To create a model with three composite $10$ representations of $SU(5)$, it
is natural to generalize the group to
$SU(5)\times Sp(m_1)\times Sp(m_2)\times Sp(m_3)\times G_D$, where
$G_D$ is a discrete symmetry.\footnote{*}{Here $Sp(n)$
is the symplectic group whose fundamental representation
is of dimension $2n$; the group
is also confusingly referred to as $Sp(2n)$.  Recall that
$Sp(1)\approx SU(2) $.}  For each choice of $m_i$ the dynamical
superpotentials and the available anomaly-free
discrete symmetries are different, and thus each case has its
own features and problems.  One reasonable choice of gauge group is
$SU(5)\times Sp(3)_1\times Sp(3)_2\times SU(2)_3\times Z_{12}$;
the first group is weakly coupled and the last three are strongly coupled,
with dynamical scales labelled $\Lambda_1$, $\Lambda_2$, $\Lambda_3$.
Define $\eta_i\equiv \Lambda_i/\mpl$. One reason for this
choice of group is that, like $SU(2)$ with six doublets,
$Sp(3)$ with ten fields $q_i$ in the fundamental representation confines
its quarks into a gauge singlet $V_{ij}=q_i q_j$ which is
antisymmetric in flavor \refs\kipp; it generates a superpotential
$W= \pf V_{ij} /\Lambda^7$, which is non-renormalizable even in terms
of the low-energy degrees of freedom.  In the present theory each $Sp(3)$
will therefore generate a $10$ and several $5$ representations of $SU(5)$
along with some singlets, but they will not have order-one
up-type Yukawa couplings.  For this gauge group there is a discrete
symmetry under which the third Higgs $H_3$ is special but under which
the $A_i$ fields have the same charge.  By using this structure we can allow
all $A_iA_jH_3$ couplings with only the top quark Yukawa coupling large.

The matter content of the model, in terms of
$SU(5)\times Sp(3)_1\times Sp(3)_2\times SU(2)_3\times Z_3\times Z_4$
multiplets, is as follows:
\eqn\modelII{\matrix{ & SU(5) & Sp(3)_1 & Sp(3)_2 & SU(2)_3 &|& Z_3 & Z_4 &\cr
\cr
x_1 & 5 & 2 & 1 & 1  &|& z^2 & -i &\cr
x_2 & 5 & 1 & 2 & 1  &|& z^2 & -i &\cr
X & 5 & 1 & 1 & 2  &|& z^2 & -i & \cr
s_1^r & 1 & 2 & 1 & 1  &|& z & i &(r=1,\dots,5)\cr
s_2^r & 1 & 1 & 2 & 1  &|& z & i& (r=1,\dots,5)\cr
S & 1 & 1 & 1 & 2  &|& z^2 & i &\cr
\bar H & \bar 5 & 1 & 1& 1  &|& z^2 & -1 &\cr
\bar Q_u & \bar 5 & 1 & 1& 1  &|& 1 & 1 &(u=1,\dots,13) \cr
\phi & 24 & 1 & 1 & 1 &|& 1 & 1 &\cr
}}
Here the discrete charges are represented by the phase
acquired by a field under the gauge transformation; $z$ is a
third root of unity.  The Planck scale superpotential can contain
the operators $\bar Q_u x_a s^r_a$, $\bar H XX\bar Q_u$,
$\bar H x_a x_a\bar Q_u$, $[X^5 S]$,   $[(x_a^2)(X^2)(XS)]$,
$[(x_a^2)(x_b^2)(XS)]$, $[(\bar  Q_u x_a s^r_a)^2]$,
$[(\bar H X S)^2]$,  {\it etc.}, and terms dependent on $\phi$.

The confining dynamics leaves us at low energies with the composite
fields $A^{[\al\bt]}_3=X^\al X^\bt/\Lambda_3$, $H^\al=X^\al S/\Lambda_3$,
and, for $a=1,2$, $r,s=1,\dots,5$, $A^{[\al\bt]}_a=x^\al_a x^\bt_a/\Lambda_a$,
$H^{\al r}_a=x^\al_a s_a^r/\Lambda_a$,  $K_a^{rs}=s_a^r s_a^s/\Lambda_a$;
here $\al,\bt$ are $SU(5)$ indices.  Their charges are
\eqn\modelIIlow{\matrix{ & SU(5)  & Z_3 & Z_4 &\cr
\cr
A_a & 10  & z & -1 & (a=1,2) \cr H_a^r & 5 & 1 & 1 & (a=1,2,\ r=1,\dots,5)\cr
K_a^{rs} & 1  & z^2 & -1& (a=1,2,\ r,s=1,\dots,5) \cr
A_3 & 10   & z  & -1\cr H & 5   & z & 1\cr
}}
These are all coupled together in the dynamical superpotential
\eqn\Wbarok{W_{dyn}= A_3A_3H + \sum_{a=1,2} {1\over \Lambda_a^2} \
\pf
\left[\matrix{A_a^{\al\bt} & H_a^{\al r} \cr
             -H_a^{\al r} & K_a^{rs} \cr} \right]
}
In addition, the superpotential at the Planck scale contributes
to the low-energy theory.
After confinement the terms $\bar Q_u x_a s_a$ generate masses of order
$\Lambda_a$ for all ten $H_a^r$, $a=1,2$, $ r=1,\dots,5$, and for
ten of the thirteen fields $\bar Q_u$.  I will refer
to the leftover fields as $\bar Q_i$, $i=1,2,3$.  The renormalizable
superpotential governing the light fields is
\eqn\GUTWc{\matrix{
W =  \sum_{i,j=1}^3 \left[ Y^{ij} A_i A_j H +
 y^{ij} A_i \bar Q_j \bar H  \right]\cr
}
}
The spectrum of couplings is now approaching that of the real world.
Recall that $\eta_i=\Lambda_i/\mpl$ and take $\eta_1<\eta_2<\eta_3$.
The coupling $Y^{33}$ is order one, so we identify $A_3$ as containing
the left-handed top and bottom quark.  The other couplings $Y^{ij}$
are naturally of order $\eta_3\eta_i\eta_j$, which establishes a
hierarchy between the top quark and the other up-type quarks.
Meanwhile, the couplings $y^{ij}$ are of order $\eta_i$.  The resulting
mass matrices, which should be compared with the matrices of Yukawa
couplings in supersymmetric models at the unification scale \refs\barger,
are therefore
\eqn\massmatII{M_u\sim\left[\matrix{
\eta_1^2\eta_3 & \eta_1\eta_2\eta_3 & \eta_1\eta_3^2\cr
\eta_1\eta_2\eta_3 & \eta_2^2\eta_3 & \eta_2\eta_3^2 \cr
\eta_1\eta_3^2 &  \eta_2\eta_3^2 & 1}\right]
\sin\bt
 \ ; \ M_d \sim \left[\matrix{
\eta_1  & \eta_1 & \eta_1\cr
 \eta_2 & \eta_2 & \eta_2\cr
\eta_3& \eta_3& \eta_3}\right]
\cos\bt
}

This spectrum is interesting, though unacceptable.  Its
general form is appealing in that it naturally implies three observed
properties of the mass matrices: the top quark is heavy, the
splittings between down-type quarks are smaller than those between
up-type quarks, and the Kobayashi-Maskawa matrix is close to unity
when the quark mass splittings are large.
However, the devil is in the details; it is not possible to get
the bottom, strange and charm quark masses in the correct ratios.
In particular, fixing the bottom and strange quark masses leaves
the charm quark too light. Even allowing that unknown coefficients of
order one might be as small as .2, no reasonable tuning brings
the masses and mixing angles into agreement with data.  One could attempt
to fix this problem by generating the charm quark mass radiatively
at low energies.  In the next two sections
I will describe two other approaches which repair the situation at the
cost of an additional parameter.

It should be noted that this mechanism does not require
$\Lambda_i>M_{GUT}$.  While for sufficiently small $\Lambda_i$ the
unification of gauge coupling constants will be disrupted, the confinement
physics will be unaffected; the weakly
coupled $SU(5)$, whether intact or broken, is a spectator to the important
dynamics.  Indeed one may give up unification and take the standard model
gauge group up to $\mpl$; the confinement physics is insensitive
to this choice.

\newsec{Improving the model}

In this section, I modify the above model slightly in order to
achieve a reasonable fermion mass spectrum.  This particular
method has the by-product that the $SU(5)$ relations between
the strange and down quark masses and those of the muon and electron
can be altered.  The trick is to adjust the
discrete symmetry so that $A_1$ and $A_2$ have opposite $Z_{12}$
charge to the choices given in \modelIIlow; I also assign $Z_3\times Z_4$
charge $(1,-1)$ to the adjoint $\phi$.
(Although it is not necessary to do so, I will also include a singlet
$S$ which has the same charge as $\phi$.)  The effect is that
the strange and down quark masses are only generated when $SU(5)$
is broken and are somewhat suppressed relative to the bottom quark mass.

Since the $A_a$, $a=1,2$, now have opposite charge to the previous
case, the coupling $A_a\bar Q_j \bar H$ is now forbidden from appearing
in the superpotential, and so the strange and down quark masses are
set to zero.  However, previously ignored operators, such as
$\bar H\phi H$ ($\bar HS H$) and
$A_a\phi\bar Q_j \bar H$ ($A_a S\bar Q_j \bar H$), can now give
masses to the Higgs bosons and to the down and strange quarks.
We may forbid the Higgs mass terms, if desired, by adding yet another
discrete symmetry; but let us keep them for the moment.  The
superpotential includes the terms
\eqn\GUTWd{\matrix{
W =  &\sum_{i,j=1}^3 Y^{ij} A_i A_j H
+ \sum_{j=1}^3y^{3j} A_3 \bar Q_j \bar H
+ \bar H(h\phi +h'S) H  \cr \cr
 &\ +{1\over\mpl} \sum_{a=1}^2 \sum_{j=1}^3
\bar H A_a \left [t^{aj}\phi + t'^{aj} S\right]\bar Q_j
}
}
The couplings $h,h'$ are of order $\eta_3$ while $t^{aj},t'^{aj}$
are of order $\eta_a$.

Let us first assume that only $\phi$ gets a vacuum expectation
value equal to $M_{GUT}\equiv \zeta\mpl$ while $\vev{S}=0$.  Then,
ignoring the fact that the Higgs bosons are given masses,
we find predictions (at $M_{GUT}$) of the following sort:
\eqn\massmatIIbI{M_u\sim\left[\matrix{
\eta_1^2\eta_3 & \eta_1\eta_2\eta_3 & \eta_1\eta_3^2\cr
\eta_1\eta_2\eta_3 & \eta_2^2\eta_3 & \eta_2\eta_3^2 \cr
\eta_1\eta_3^2 &  \eta_2\eta_3^2 & 1}\right]
\sin\bt
 \ ; \ M_d \sim \left[\matrix{
\eta_1 \zeta & \eta_1\zeta & \eta_1\zeta\cr
 \eta_2\zeta & \eta_2 \zeta& \eta_2\zeta\cr
\eta_3& \eta_3& \eta_3}\right]
\cos\bt
}
Again the Kobayashi-Maskawa matrix is nearly diagonal as a result
of the confinement hierarchy and the discrete symmetries.
Notice that to get the down quark masses in the correct proportion
while maintaining a reasonable charm quark mass we
actually need $\eta_2>\eta_3>\eta_1$ for small $\tan\beta$,
while for large $\tan\beta$ the hierarchy requires $\eta_3>\eta_2>\eta_1$.
It appears that $\tan\bt\sim 1$ cannot be
accommodated.\footnote{*}{If the Higgs boson $\bar H$ or the fields
$\bar Q_i$ are also composite, then
additional suppression factors will reduce the entire down-quark
matrix uniformly, allowing smaller values for $\tan\bt$, and having
no easily observable effect at low energy.}  Also, $\zeta$ cannot be too
small without driving $\eta_2$ close to one, at which point a field
theoretic discussion of confinement breaks down.
The mixing angles are of the right order of magnitude; the Cabbibo angle tends
to be too small but is the most sensitive of the angles to the
specific values of the coefficients in the two mass matrices.

Note also that the standard $SU(5)$ relations for lepton and down-quark
masses have been altered by this mechanism, though the direction
of the effect should be the same for both generations, while in fact
the ratios $m_s/m_d$ and $m_\mu/m_e$ are far from equal.
When the singlet $S$ also acquires a vacuum expectation value, we have a
hope of killing two quarks with one stone.  Suppose that a version
of the sliding singlet mechanism \refs\sswitt\
could be used here, solving the doublet-triplet splitting problem
by making $\vev{S}$ and $\vev\phi$ proportional.
(Recall that such a mechanism can be stable if supersymmetry breaking
occurs at a low scale \refs\ssnem.)
Simultaneously, if the coefficients $t^{aj},t'^{aj}$ have no particular
symmetry, the $SU(5)$ relations for the two light generations
would be broken, and even the ratios $m_s/m_d$ and
$m_\mu/m_e$ would be unrelated to one another.

\newsec{A second model with an acceptable spectrum}

Another way to build a theory which can lead to acceptable fermion
masses is to restrict the couplings $y$ by a symmetry so that the
down-type Higgs boson couples only to $A_3$ at leading order.
Let us return to the gauge group $SU(5)\times [SU(2)]^3\times Z_{12}$.
Under the $Z_3\times Z_4\approx Z_{12}$, the fields $X_3, S_3, \bar H$ are
assigned charge $(z^2,-i),\ (z^2,i),\ (z^2,-1)$ while $X_a, S_a$ are
assigned charge $(1,i),\ (1,-i)$ for
$a=1,2$; another five $\bar 5$ representations $\bar Q_u$ are neutral
under $Z_{12}$. After confinement, $H_1$ and $H_2$ are neutral under the
discrete symmetry and become massive along with two of the $\bar Q_u$; label
the remaining three fields $\bar Q_i$, and relabel $H\equiv H_3$.
The fields $A_1, A_2$ have discrete charge $(1,-1)$, while
the charges for $A_3,\bar H, H$ are $(z,-1),\ (z^2,-1),\ (z,1)$.
The renormalizable terms in the superpotential for the massless fields are
\eqn\GUTWb{\matrix{
W =    Y A_3 A_3 H
+ y A_3 \bar H \bar Q_3 \cr
}
}
where the flavor index of the $\bar Q_i$ was rotated so that only
$\bar Q_3$ appears in the above formula.  This model is just the one-generation
model we started with, plus two massless generations.

Giving masses to the other quarks and leptons requires
partially breaking the discrete symmetry.  The operators
$A_a\bar Q_j\bar H$ and $A_aA_b H$, where $a,b=1,2$ and $j=1,2,3$,
have charge $(z^2,1)$ and $(z,1)$ respectively under the $Z_3\times Z_4$.
If a gauge singlet
$S$ with discrete charge $(z^2,1)$ acquires a vacuum expectation value
$\vev{S}=\xi\mpl$, then it will allow light quark masses to be generated.
In particular the charm and up quark masses will be suppressed by $\xi$
and those of the strange and down quarks will be suppressed by $\xi^2$.
Dangerous terms like $\bar H H$ and $A_i\bar Q_j\bar Q_k$ and allowed terms
of the form $A_3 A_a H$ cannot be generated, since their charges are
not multiples of $(z,1)$. Unfortunately the term $\bar Q_i H$ {\it can}
be generated; this can only be forbidden by adding yet another discrete
symmetry (R-parity) under which $A_i,\bar Q_i$ change sign while
$H,\bar H$ do not.

The $SU(2)$ groups for the three generations become strongly coupled
at scales $\Lambda_1$, $\Lambda_2$, $\Lambda_3$;
define $\eta_i\equiv\Lambda_i/\mpl$.
The high-energy mass matrices, up to factors of order one, are then
\eqn\massmatI{M_u\sim\left[\matrix{
\eta_1^2\eta_3\xi & \eta_1\eta_2\eta_3\xi & \eta_1\eta_3^2\xi^2 \cr
\eta_1\eta_2\eta_3\xi & \eta_2^2\eta_3\xi & \eta_2\eta_3^2\xi^2 \cr
\eta_1\eta_3^2\xi^2&  \eta_2\eta_3^2\xi^2 & 1}\right]
\sin\bt
\ ; \ M_d \sim \left[\matrix{
\eta_1 \xi^2 & \eta_1\xi^2 &   \eta_1\xi^2\cr
 \eta_2\xi^2 & \eta_2\xi^2 &
\eta_2\xi^2\cr \eta_3& \eta_3& \eta_3}\right]
\cos\bt
}
Again the Kobayashi-Maskawa matrix will be close to diagonal,
as has been guaranteed by the hierarchy in the confinement scales
and the discrete symmetry.  For this model $\tan\bt$ must be large,
with $\eta_2\sim\eta_3>\eta_1$ and $\xi\sim .2$.  (As before,
$\tan\bt$ can be smaller if $\bar Q_i$ or $\bar H$ are also composite.)
Again the mixing angles are of the right
order of magnitude, with the Cabibbo angle tending to be too
small but varying rapidly as coefficients are adjusted.

\newsec{Summary}

While these models in their present form win no prizes for elegance,
do not by themselves break $SU(5)$ and supersymmetry, and do not
consistently generate a small $\mu$ term and a large mass for the
color-triplet Higgs bosons,  they have a number of interesting features
which can perhaps be used in more complete and successful models.

(1) The generations of the standard emerge in a curious way.
Those particles which are contained in $\bar 5$ representations
of $SU(5)$ -- the down-type antiquarks and the down-type Higgs
and lepton doublets -- are present as fundamental
fields, while all other particles, in the $5$ and $10$ representations,
arise as massless composites below the scales $\Lambda_i$ of the
confining gauge groups.

(2) The confining dynamics has implications for the masses
of the quarks. The top quark gets its
mass from an operator which is generated dynamically during
confinement; its coefficient is of order one.
 The other quarks and the leptons get their
masses from operators which at $\mpl$ have dimension at least four in
the superpotential (five in the Lagrangian);
their Yukawa couplings to the Higgs bosons are of
order $\Lambda/\mpl$ to a positive power.
A hierarchy of confinement scales is inherited by the
Yukawa couplings, ensuring that the Kobayashi-Maskawa
matrix is close to the unit matrix when the quark masses
are very different from one another.  Furthermore, the tendency is
for the splittings of masses of adjacent generations to be larger in
the up-quark sector than in the down-quark sector.

(3) In the simplest illustration of the mechanism, the spectrum of
predicted quark mass relations is difficult
to reconcile with the observed masses.  Two models are proposed in which
certain quark masses and mixings are only generated when a discrete symmetry
is broken at a lower scale; this introduces a new parameter into the
theory and improves the spectrum at the cost of predictivity.  In one version
the strange and down quark masses are generated during
$SU(5)$ breaking, potentially destroying their relations
with lepton masses.

(4) In these models, the predictions of $SU(5)$ grand unification
are naturally preserved, since the $SU(5)$ group is a weakly coupled
spectator to the dramatic events of confinement.  This is true even when
$M_{GUT}$ lies at or somewhat above the confinement scale(s).
If one gives up on $SU(5)$ unification the mechanism will
still work with the standard model gauge group. Since the left-handed
down quark is composite while the right-handed down-quark is not,
this mechanism cannot be directly transplanted to models with
$SO(10)$ or $E_6$ unified gauge groups.

  Finally, it is worth commenting on the remarkable fact that the
analysis of this scenario is firmly based on developments in our
understanding of strongly coupled supersymmetric gauge theories.
Most composite models have relied on questionable dynamical assumptions.
The dynamics discussed here are known to be consistent with a wide
variety of phenomena found in supersymmetric gauge
theories \refs{\kinsrev}.  Note, in particular, that the choice
of gauge group $SU(2)$ was essential.  For example, if $[SU(2)]^3$
were simply replaced with $[Sp(2)]^3$\footnote{*}{$Sp(2)$ is the
symplectic group with a four-dimensional fundamental representation;
it is often called $Sp(4)$.}, the
low energy $SU(5)$ composite representations would still be
$A_i$ and $H_i$, but no top-quark--Higgs-boson Yukawa coupling would
be generated,  and a quantum mechanical constraint
$\vev{A^2H}=\Lambda^3$ would break $SU(5)$ to at most
$SU(2)\times SU(2)$ at the confinement scale \refs{\seiberg}! Adding two
additional fundamental representations of each $Sp(2)$ to the model
would leave $SU(5)$ unbroken but still would not lead to a
top-quark--Higgs boson Yukawa coupling.
Larger groups would be even more unstable to symmetry breaking.
Thus, the dynamics of this model is quite special.

  In summary, the dynamics of confinement as understood in \refs\seiberg\
have been applied to an extension of the minimal supersymmetric
standard model, with the fields and couplings of the standard model
emerging only at low energy.   The successful predictions of $SU(5)$
grand unification for the gauge couplings and the tau-lepton and
bottom-quark Yukawa couplings
can be preserved despite the strong coupling phenomena.
The mass hierarchies and diagonal Kobayashi-Maskawa matrix
are explained as due to a hierarchy in the confinement scales of
the three generations in conjunction with a discrete symmetry.
The large top quark mass and the larger mass splittings in the
up-quark sector versus the down-quark sector are natural predictions
of the mechanism, though to get the bottom, charm and strange quark masses
to be consistent apparently requires fine tuning or additional
structure.  Several implementations of this mechanism
with a minimal number of Higgs bosons have been presented.   One simple model
generates a spectrum which has many good features but is probably
ruled out; two other variants give reasonable fermion mass spectra, though
both are complicated and incomplete.
Many other variants are possible, so perhaps more successful and
elegant models using this mechanism can be found, or perhaps
other theories can be invented which contain the special features of
this scenario.  Even should it prove to be a dead end, this
work demonstrates that our improved understanding
of gauge theories makes it possible to build strongly coupled models
which have interesting dynamics, can be analyzed reliably, and resemble
the real world.

\bigskip

\centerline{{\bf Acknowledgments}}

It is a pleasure to thank the faculty, postdocs and visitors
at Rutgers University for useful discussions and comments.  I am
particularly indebted to L.~Ibanez and R.~Ratazzi.  I am also
grateful to A.~Nelson with whom I had conversations on related
topics.  This work was supported in part by DOE grant \#DE-FG05-90ER40559.

\listrefs
\end